\documentclass[aps,prl,twocolumn,groupedaddress,showpacs]{revtex4}

\usepackage{graphicx}
\usepackage{amsmath}
\usepackage{bm}     

\newcommand{\ban}{\begin{eqnarray*}}
\newcommand{\ean}{\end{eqnarray*}}
\newcommand{\bea}{\begin{eqnarray}}
\newcommand{\eea}{\end{eqnarray}}

\newcommand{\bra}[1]{\left\langle #1 \right|}
\newcommand{\ket}[1]{\left|#1\right\rangle}

\newcommand{\B}[1]{\mathbf{#1}}

\newcommand{\Hprime}{\mathcal{H^\prime}}
\newcommand{\norm}[1]{\left\vert #1\right\vert}
\newcommand\Z{\mathrm{Z}}
\newcommand\Bpar{\ensuremath{\B{B}_\parallel}}
\newcommand\Hso{\mathcal{H}^\mathrm{so}}
\newcommand\Hsosfdag{{\mathcal{H}^\mathrm{so}_{\text{SF}}}^\dagger}
\newcommand\Hdsf{\mathcal{H}^\text{c}_\text{SF}}
\newcommand\Hp{\mathcal{H}^\prime}
\newcommand\HRMT{\ensuremath{\mathcal{H}_\text{RMT}^\prime}}
\newcommand\eVA{eV$\text\AA^3$}
\newcommand\Aso{{\bf A}_\text{so}}
\newcommand\LamD{\ensuremath{\Lambda_\text{c}}}
\newcommand\etal{\emph{et al.\ }}
\newcommand\etall{\emph{et al}}
\newcommand\apcubic{\ensuremath{a_{\Vert,\text{c}}}}
\newcommand\aplinear{\ensuremath{a_{\Vert,\text{l}}}}
\newcommand\avg{\ensuremath{\langle g \rangle}}

\newcommand{\abs}[1]{\left\lvert#1\right\rvert}

\begin{document}


\title{Cubic Dresselhaus Spin-Orbit Coupling in 2D Electron Quantum Dots}


\author{Jacob J.\ Krich}
\author{Bertrand I.\ Halperin}
\affiliation{Physics Department, Harvard University, Cambridge, MA
02138}


\date{\today}

\begin{abstract}
We study effects of the oft-neglected cubic Dresselhaus spin-orbit
coupling (i.e., $\propto p^3$) in GaAs/AlGaAs quantum dots. Using a
semiclassical billiard model, we estimate the magnitude of the
spin-orbit induced avoided crossings in a closed quantum dot in a
Zeeman field. Using these results, together with previous analyses
based on random matrix theory, we calculate corresponding effects on
the conductance through an open quantum dot. Combining our results
with an experiment on conductance through an 8 $\mu$m$^2$ quantum
dot [D.\ M.\ Zumb\"uhl \emph{et al}., Phys.\ Rev.\ B \textbf{72},
081305 (2005)] suggests that 1) the GaAs Dresselhaus coupling
constant, $\gamma$, is approximately 9 \eVA, significantly less than
the commonly cited value of 27.5 \eVA and 2) the majority of the
spin-flip component of spin-orbit coupling can come from the cubic
Dresselhaus term.
\end{abstract}



\maketitle



Control over electron spin in semiconductors has promise for quantum
computing and spintronics.  In such applications, it is essential to
understand how the transport of an electron through a circuit
affects its spin; i.e., we must understand spin-orbit coupling. In
technologically important III/V semiconductor heterostructures,
spin-orbit coupling originates in the asymmetry of the confining
potential (called the Rashba term), which can be controlled by
gates, and in bulk inversion asymmetry of the crystal lattice
(called the Dresselhaus term). In quasi-2D systems, the Dresselhaus
term has two components, one linear in the electron momentum and the
other cubic. The cubic Dresselhaus term is usually neglected, as it
is generally smaller than the linear contribution. Datta and Das
proposed a spin-field-effect transistor (SFET) for quasi-1D
ballistic wires with Rashba coupling \cite{datta90}. Schliemann,
Egues, and Loss proposed an SFET that can operate in diffusive
quasi-2D systems based on tuning the Rashba and linear Dresselhaus
terms to be equal in strength, which produces long spin lifetimes,
neglecting the cubic Dresselhaus term \cite{schliemann03}. The
strengths of the spin-orbit terms are difficult to measure
independently, but a full understanding of their strengths is
crucial to making such devices. Additionally, in confined systems
such as quantum dots, some effects of the linear spin-orbit terms
are suppressed \cite{halperin01}, and it is important to know the
magnitude of the cubic Dresselhaus contribution, which could limit
or even prevent the functioning of spintronic devices.

We characterize the strength of the cubic Dresselhaus term in a
confined system by its effect on avoided crossings in an in-plane
magnetic field \Bpar{} that couples only to the electron spin. With
no spin-orbit coupling, each eigenstate can be written as a product
of orbital state $\ket\alpha$ and spin quantized along \Bpar.
Eigenstates $\ket{\alpha \uparrow}$ and $\ket{\beta \downarrow}$
become degenerate when $\epsilon_\alpha-\epsilon_\beta=E_\Z$, where
$\epsilon_{\alpha,\beta}$ are the orbital energies and $E_\Z$ is the
Zeeman energy \footnote{This neglects other spin-flip processes,
e.g., hyperfine coupling, which in GaAs is much smaller than
spin-orbit effects for quantum dots on the micron scale.}, but
spin-orbit coupling leads to avoided crossings.

In the first half of this Letter, we estimate the cubic Dresselhaus
contribution to the avoided crossings, which can be larger than the
linear terms' contribution since the latter are suppressed for small
\Bpar{} \cite{halperin01}. In the second half of this Letter, we
relate these avoided crossings in closed quantum dots to the mean
and variance of the conductance when the quantum dot is connected to
ideal leads. We compare these predictions to the results of
Zumb\"uhl \etal on transport through an 8 $\mu$m$^2$ quantum dot
\cite{zumbuhl05,zumbuhl02} and find that agreement
is possible only if the cubic Dresselhaus coupling constant in GaAs,
$\gamma$, is considerably less than the frequently cited value of
27.5~\eVA{} \cite{knap96,winkler03} from $\B{k}\cdot\B{p}$ theory. A
smaller value of $\gamma$ has also been suggested by experiments
\cite{richards96,jusserand95} and band structure studies
\cite{santos94,santos95,chantis06}. Even with this smaller value of
$\gamma$, we find that the cubic Dresselhaus term is the dominant
spin-flip mechanism in the sample considered. Both of these results
are significant for spintronic devices.

We consider conduction electrons in a 2D electron system grown on a
(001) surface of a III/V semiconductor confined to a small area by a
potential $V(\B{r})$. We use an effective Hamiltonian
\begin{eqnarray}\label{eq:rotatedhamiltonian}
\begin{split}
  \mathcal{H}=\frac{\left({\bf p}-\Aso\right)^2}{2m}
  &+\frac{\gamma}{2\hbar^3}(p_2^2-p_1^2)(\B{p}\times{\bm{\sigma}})\cdot\hat{\B{e}}_3\\
  &+V(\mathbf{r}) +\frac{1}{2}g\mu_B\mathbf{B}\cdot\bm{\sigma}
\end{split}
\end{eqnarray}
where $\B{p}=\B P-e \B A/c$, $\B{P}$ is the canonical momentum,
$\B{A}$ is the vector potential from the perpendicular magnetic
field, $\bm\sigma$ is the vector of Pauli matrices, $m$ is the
effective mass, $\Aso={\bf\hat{e}}_1\hbar\sigma_2/2\lambda_1
-{\bf\hat e}_2\hbar\sigma_1/2\lambda_2$ is the effective spin-orbit
vector potential, which contains both the Rashba and linear
Dresselhaus spin-orbit terms, and $\lambda_{1,2}$ are the (linear)
spin-orbit lengths \cite{aleiner01,cremers03}. We choose a
coordinate system with axes $\mathbf{\hat e_1}$=[110], $\mathbf{\hat
e_2}$=[1$\bar{1}$0], and $\mathbf{\hat e_3}$=[00$\bar1$]. The second
term is the cubic Dresselhaus term.

In a system of linear size $L$, the linear spin-orbit terms can be
gauged away to first order in $L/\lambda$ by the unitary
transformation $\mathcal{H}\rightarrow
U\mathcal{H}U^\dagger\equiv\mathcal{H^\prime}$ where $U=\exp(i\bf r
\cdot \Aso)$ \cite{aleiner01}. Expanding to leading
order in $L/\lambda$, %
\bea %
    \Hprime&=&\frac{1}{2m}(\B{p}-\B{a_\bot}-\B{a}_\Vert)^2
    +b^\Z +b_\bot^\Z
    \nonumber\\
    &+&\frac{\gamma}{2\hbar^3}(p_2^2-p_1^2) 
    (\B{p}\times{\bm{\sigma}})\cdot\hat{\B{e}}_3
    +V(\B{r}) \label{eq:expandedhamiltonian},
\eea %
 where
$\B{a}_\bot=\frac{\hbar\sigma_3}{4\lambda_1\lambda_2}[\B{\hat{e}}_3\times\B{r}]$,
$\B{a}_\Vert=\frac{\hbar}{6\lambda_1\lambda_2}(x_1\sigma_1/\lambda_1
+x_2\sigma_2/\lambda_2)[\B{\hat{e}}_3\times\B{r}]$, $b^\Z=g\mu_B
\B{B}\cdot \bm{\sigma}/2,$ and $b_\bot^\Z=-g\mu_B(B_1
x_1/\lambda_1+B_2 x_2/\lambda_2)\sigma_3/4$. %

When we apply a Zeeman field, we can treat each induced degeneracy
as a two-level system, assuming the spin-orbit matrix elements,
$\epsilon_\text{so}$, are much less than the single spin mean level
spacing, $\Delta=2\pi\hbar^2/m A$, with $m$ the conduction band
effective mass and $A$ the dot area. The magnitude of the avoided
crossings at the Fermi energy is given by
$\epsilon_\mathrm{so}=\left\vert
\bra{\alpha\uparrow}\Hso\ket{\beta\downarrow}\right\vert$, where
$\epsilon_\alpha-\epsilon_\beta=E_\Z$ and $\epsilon_\alpha=E_F$. We
want to find the rms value of $\epsilon_\mathrm{so}$. Following
Ref.\ \cite{halperin01}, for a closed chaotic dot we may write
$\Lambda^2\equiv\overline{(\epsilon_\text{so}/\Delta)^2}$ as%
\bea
    \Lambda^2
    =\overline{\sum_{\alpha \beta}\norm{(\Hso)_{\alpha\uparrow,\beta
    \downarrow}}^2 \delta(\epsilon_\alpha-\epsilon_\beta-E_Z)
    \delta(\epsilon_\alpha-E_F)}, \label{eq:lambdadelta}
\eea %
where the overbar indicates ensemble averaging and
$O_{a,b}\equiv\bra{a}O\ket{b}$.

We rewrite Eq.\ \ref{eq:lambdadelta} as in Ref.\ \cite{halperin01}
using the $t$-dependent representation of the delta function and
interaction picture operators, and, after summing over $\beta$, find
\bea%
    \Lambda^2=\int_{-\infty}^\infty \frac{\text{d}t}{\Delta2\pi\hbar}
    e^{-i\omega_Z t} \overline{\bra{\alpha}\Hso_\text{SF}(t)
    \Hsosfdag(0)\ket\alpha} \label{eq:lambda},%
\eea %
where $\omega_Z=E_\mathrm{Z}/\hbar$ is the Zeeman frequency,
$\ket\alpha$ is a typical orbital eigenstate with
$\epsilon_\alpha\approx E_\text{F}$, and
$\Hso_\mathrm{SF}\equiv\bra{\uparrow}\Hso\ket{\downarrow}$ is the
spin-flip part of $\Hso$.  We consider the cubic Dresselhaus term
alone, $\mathcal{H}^{c}=\frac{\gamma}{2\hbar^3}
(p_2^2-p_1^2)(p_1\sigma_2-p_2\sigma_1)$, and estimate its
contribution to $\Lambda$, which we call $\LamD$.  For example, if
$\Bpar=B\hat{\B{e}}_1$,
$\mathcal{H}^{c}_\mathrm{SF}=\frac{\gamma}{2\hbar^3}(p_2^2-p_1^2)p_1$.

We estimate $\LamD$ semiclassically using a billiard model for the
quantum dot, where the matrix element in Eq.\ \ref{eq:lambda} is
replaced by the corresponding expectation value for a classical
particle moving at the Fermi velocity, $v_F$, starting at a random
point moving in a random direction. We consider $B_\bot=0$ for these
simulations. Each of 2-3$\times10^5$ such trajectories is followed
for an equal amount of time, which is generally about 300 bounces
total in the forward and backward directions. Increasing the number
of trajectories or bounces does not change the results. We calculate
$\int \text{d}t e^{-i\omega_Z (t-t^\prime)}\Hdsf(t)\Hdsf(t^\prime)$
for 100 random initial times $t^\prime$ on each trajectory as a
function of $\omega_Z$, and their average gives $\LamD^2$ when
multiplied by the appropriate prefactors. We add a damping function
to the integrand that sends it smoothly to zero as $t$ approaches
the simulation cutoff.
\begin{figure}  
    \includegraphics[width=3.375in]{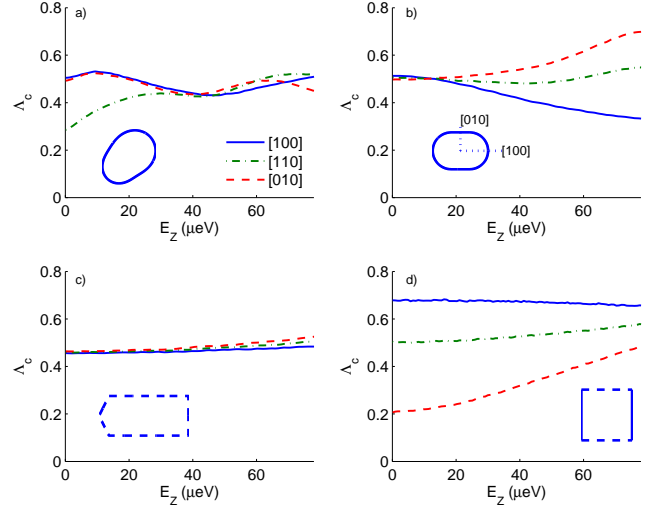}%
    \caption{(color online) Normalized rms avoided crossing due to
    cubic Dresselhaus spin-orbit coupling, $\LamD$, as a function of
    Zeeman energy $E_Z$ for four billiards with in-plane magnetic
    field along the indicated directions, with $\gamma=8.5$~\eVA
    \cite{chantis06}; all billiards have $A=8$~$\mu$m$^2$ and
    $n=5.8\times 10^{15}$~m$^{-2}$.  Insets show the billiard shapes
    with crystal axes. Solid lines indicate specular boundary
    conditions and dashed lines indicate diffuse boundary
    conditions. a) has a mixed phase space with small regions of
    regular trajectories, b) is a stadium billiard, c)
    is similar to the dot in Ref.\ \cite{zumbuhl05}, with
    diffuse boundaries to ensure chaos, and d) is a square
    with diffuse scattering from the top and bottom and specular
    scattering from the sides (see text).}
    \label{fig:billiardlambda}
\end{figure}

We consider four different billiard shapes and, for specificity,
choose parameters corresponding to the largest, highest density
quantum dot in Ref.\ \cite{zumbuhl05}, with $A=8$ $\mu$m$^2$ and
$n=5.8\times10^{15}$ m$^{-2}$.  We use $g=0.44$ and
$m=0.067m_\text{e}$ where $m_\text{e}$ is the electron mass. Fig.\
\ref{fig:billiardlambda}(a-c) shows the resulting $\LamD(E_Z)$ for
these billiards, for three orientations of $\Bpar$, with the
Dresselhaus coupling constant chosen to be $\gamma=8.5$ \eVA. For
other choices, $\LamD$ scales linearly with $\gamma$. The insets
show the shapes of the billiards. For the semiclassical method of
estimating $\LamD$ to be valid, we must have $\Delta\ll E_Z \ll
E_T$, where $E_T=\hbar v_F/\sqrt{A}$ is the Thouless energy. For the
case discussed here, $\Delta=0.9$ $\mu$eV and $E_T\approx80$
$\mu$eV.

We can understand the approximate scale of $\LamD$ by using a
simpler, unphysical billiard. Consider an $L_x\times L_y$ rectangle
with specular reflection boundary conditions on the sides and
diffuse boundary conditions on the top and bottom. At each collision
with a diffuse wall, we choose the tangential momentum from a
uniform distribution on [-$p_F$, $p_F$] \cite{Fliesser96}. This
choice gives a correct weighting for diffuse scattering and
maintains detailed balance. In such a billiard, for $\Bpar
\parallel \hat{\B{x}}$,
in the limit $E_Z\rightarrow0$,%
\bea%
    \LamD^2\Bigr\vert_{E_Z\rightarrow0}=\frac{\gamma^2}{\Delta
    2\pi\hbar^7}\left\langle\int_{-\infty}^{\infty} \text{d}t p_y
    p_x^2(t)p_y p_x^2(0)\right\rangle \label{eq:lambdaestimate},
\eea %
and we can break each trajectory into segments between collisions
with the top/bottom walls.  Along each segment, $p_x^2$ and $p_y$
are constant, and the particle takes time $t=m L_y/\abs{p_y}$ to
move from one end of the segment to the other, so we can rewrite
Eq.\ \ref{eq:lambdaestimate} as%
\bea
    \LamD^2\Bigr\vert_{E_Z\rightarrow0}=\frac{\gamma^2 m L_y}{\Delta
    2\pi\hbar^7} \left\langle
    \abs{p_y(0)} p_x^2(0)
    \sum_{n=-\infty}^\infty (-1)^n p_{x,n}^2\right\rangle,
\eea%
which we can evaluate explicitly, since the $p_{x,i}^2$ are
uncorrelated between segments. The particle begins moving in a
random direction with $\mathcal P(p_{x,0})=\pi^{-1}
(p_F^2-p_{x,0}^2)^{-1/2}$, where $\mathcal P$ is the probability
density on $[-p_F$, $p_F]$. Since the diffuse boundaries in this
billiard are the top and bottom, $\mathcal P(p_{x,i\ne 0})=1/2p_F$.
We regularize the infinite sum by $\sum_{n=-\infty}^\infty(-1)^n=0$,
and, noting that $E_T=\hbar p_F/m L_y$, we find $\LamD^2=4\gamma^2
p_F^6/(45\pi^2 \Delta E_T \hbar^6)$. For the parameters in Fig.\
\ref{fig:billiardlambda}, this gives $\LamD(E_Z\rightarrow0)=0.678$.
Finite values of $E_Z$ are not amenable to such simple treatment,
but simulations of this billiard appear in
Fig.~\ref{fig:billiardlambda}d, where the results for $\Bpar
\parallel \hat{\mathbf{x}}$, shown by the solid trace, approach the analytic
prediction for $E_Z\rightarrow0$.

Avoided crossings have not yet been directly measured in chaotic
dots, but our calculations can be related to experiments measuring
the conductance $g$ through a quantum dot by Zumb\"uhl \etal
\cite{zumbuhl02,zumbuhl05}. To make this comparison, we need a
connection between avoided crossings in a closed dot and properties
of the dot with leads attached. Cremers \emph{et al.}, using random
matrix theory (RMT), worked out a similar connection for dots with
only Rashba and linear Dresselhaus spin-orbit coupling
\cite{cremers03}. We point out that the cubic Dresselhaus term can
be added easily into the predictions of Cremers \etal \emph{without}
changing their formulas for $\langle g\rangle$ and var~$g$ by
reinterpreting one of their RMT energy scales to include both linear
and cubic spin-orbit terms. We now elaborate.

In Ref.~\cite{cremers03}, the chaotic quantum dot is connected to
two ideal leads with $N\gg1$ open channels, giving a scattering
matrix from the circular orthogonal ensemble. They treat the
magnetic field and spin-orbit coupling with a stub model
\cite{brouwer97} in which the stub has the $M\times M$ perturbation
Hamiltonian, $\HRMT$, given by association to $\Hp$ in Eq.\
\ref{eq:expandedhamiltonian} (without
the cubic Dresselhaus term), as%
\bea
    \HRMT &=&\frac{\Delta }{2\pi }
    \Bigr[i\mathcal{A}_0(x\openone+a_{\bot }\sigma_{3})
    +ia_\Vert\left(\mathcal{A}_{1}\sigma_{1}
    +\mathcal{A}_{2}\sigma_{2}\right)\nonumber\\
    &&-\mathbf{b}\cdot \mbox{\boldmath $\sigma$}
    +b_{\bot}\mathcal{B}_{h}\sigma_{3})\Bigr] \label{eq:RMT_H},
\eea %
where $\mathcal{A}_i$, $i=0,1,2$, are real antisymmetric matrices
with $\left\langle \mbox{tr}\, \mathcal{A}_{i}
\mathcal{A}_{j}^{T}\right\rangle =\delta_{ij}M^{2}$,
$\mathcal{B}_{h}$ is a real symmetric matrix with $\left\langle
\mbox{tr}\,\mathcal{B} _{h}^{2}\right\rangle =M^{2}$, $M\gg1$ is the
number of channels in the stub, and $x$, $a_\bot$, $a_\Vert$, $\B
b$, and $b_\bot$ are dimensionless parameters, with $x$
corresponding to $\B B_\bot$, $\B b$ to the Zeeman field, and
$a_\bot$, $a_\Vert$, and $b_\bot$ to the similarly named terms in
Eq.~\ref{eq:expandedhamiltonian} (without the cubic Dresselhaus
term). Dephasing is included by setting
$N_\text{eff}=N+2\pi\hbar/\tau_\phi\Delta$, where $\tau_\phi$ is the
dephasing time. Expressions are then obtained for $\langle g\rangle$
and var~$g$ as functions of $x$, $a_\bot$, $a_\Vert$, $\B b$,
$b_\bot$, and $N_\text{eff}$ to leading order in $1/N_\text{eff}$
\cite{cremers03}. Zumb\"uhl \etal use these results to fit their
data.

Without the cubic Dresselhaus term, the correspondence between Eqs.\
\ref{eq:RMT_H} and \ref{eq:expandedhamiltonian} gives the following
mapping from physical parameters to RMT parameters:
\begin{eqnarray}\label{eq:energyscales}
  x^2&=&\pi \kappa E_T/\Delta (\Phi/\Phi_0)^2 \quad\quad\quad
  b=\pi E_Z/\Delta\nonumber \\
  a_\bot^2&=&\pi\kappa E_T/\Delta (A/4\lambda_{SO}^2)^2 \\
  a_\Vert^2&=&\pi\kappa^\prime E_T/\Delta
  (A/4\lambda_{SO}^2)^2((L_1/\lambda_1)^2+(L_2/\lambda_2)^2) \nonumber\\
  b_\bot^2&=&\pi \kappa^{\prime\prime} (E_Z)^2/E_T\Delta
  (A/4\lambda_{SO}^2)\nonumber
\end{eqnarray}
where $\Phi$ is the magnetic flux through the quantum dot,
$\Phi_0=h/2e$ is the flux quantum,
$\lambda_{SO}=\sqrt{\lambda_1\lambda_2}$, $L_{1,2}$ are the linear
dimensions of the roughly rectangular dot, oriented along
$\mathbf{\hat e}_{1,2}$, and $\kappa$, $\kappa^\prime$, and
$\kappa^{\prime\prime}$ are geometric factors of order unity
\cite{cremers03,aleiner01}.
We add the cubic Dresselhaus to this theory by noting that, as a
random matrix, the cubic Dresselhaus term in
Eq.~\ref{eq:expandedhamiltonian} has the same symmetry as the
$a_\Vert$ term in Eq.~\ref{eq:RMT_H}, i.e., it contains only
$\sigma_1$ and $\sigma_2$ Pauli matrices. Assuming no correlation
with the linear terms, we include the cubic Dresselhaus in \HRMT{}
by setting $a_\Vert^2=\aplinear^2+\apcubic^2$, where \aplinear{} is
the Rashba and linear Dresselhaus contribution, given by
Eq.~\ref{eq:energyscales}, and \apcubic{} is the cubic Dresselhaus
contribution. Since \HRMT{} contains the spin-orbit part of the
Hamiltonian of the closed quantum dot, we relate \apcubic{} to
\LamD{} by finding the rms spin-flip matrix element (with spins
quantized along \Bpar{}) due to \apcubic{}, giving
$\apcubic=2\pi\LamD$. Including the cubic Dresselhaus term in
\HRMT{} in this way cannot improve any fits to experimental data, as
it does not change the fitting equations. However, its inclusion
lifts the constraint that $a_\Vert\ll a_\bot$ \cite{aleiner01},
similar to spatially varying spin-orbit strengths \cite{brouwer02}.

Zumb\"uhl \etal observe weak anti-localization (WAL) in only one of
the GaAs/AlGaAs heterostructure quantum dots they study
\cite{zumbuhl02,zumbuhl05}, and that dot gives the best defined
values of the RMT parameters; we use it for the discussion of our
results. The other dots do not contradict this discussion. The dot
that displays WAL has area $A=8$~$\mu$m$^2$ and electron density
$n=5.8\times10^{15}$~m$^{-2}$, which are the values used in Fig.\
\ref{fig:billiardlambda}. The dot has $N=2$ and $N_\text{eff}=13.9$
\cite{zumbuhl05}.

Zumb\"uhl \etal{} measure var~$g$ as a function of \Bpar{} with time
reversal symmetry broken by a small $B_\bot$. They fit to the
expression of Cremers \etall. \cite{cremers03}, with $a_\Vert$ (and
all parameters except $\kappa^{\prime\prime}$) fixed to the value
determined from the $\langle g \rangle$ data. We redo the fits to
the var~$g$ data, constraining only $a_\Vert\ge\apcubic{}$, with
\apcubic{} from our simulations, and $\tau_\phi$ fixed to the value
determined from $\langle g \rangle$. From Fig.\
\ref{fig:billiardlambda}, a typical value of $\LamD$ in all our
billiard shapes is $0.4$, giving $\apcubic=2.5$ $(8.1)$ for
$\gamma=8.5$ $(27.5)$ \eVA  (recalling that $\LamD \propto\gamma$).
We find that a value of $\apcubic\approx 2.5$ is compatible with the
experimental data.  However, if we require that $a_\Vert\ge8.1$, the
fits to the data become markedly worse.

Zumb\"uhl \etal{} also measure $\langle g\rangle$ as a function of
$B_\bot$, which they use to determine $a_\Vert$, finding
$a_\Vert=3.1$ \cite{zumbuhl02,zumbuhl05,zumbuhlPC}. Since
$a_\Vert^2=\aplinear^2+\apcubic^2$, we must have
$a_\Vert\ge\apcubic$, so we conclude that $\gamma=27.5$ \eVA{} is
inconsistent with these results, while $\gamma=8.5$ \eVA{} is
consistent with the data. So both $\langle g \rangle$ and var~$g$
data indicate that $\gamma$ should be closer to 9 \eVA than 28 \eVA
\footnote{Even using $\LamD=0.2$, the lowest value in Fig.\
\ref{fig:billiardlambda}, gives $\apcubic=4.1$ for the larger value
of $\gamma$, which is \emph{still} inconsistent with the
experiment.}. Moreover, even with the smaller value of $\gamma$, the
cubic Dresselhaus term gives the dominant contribution to $a_\Vert$.

There are only a few other experiments pertaining to the value of
$\gamma$ in GaAs.  The best, most direct study is the Raman
scattering in a GaAs/AlGaAs quantum well by Richards \etal in which
they found $\gamma=11.0$~\eVA{} \cite{richards96}; the same group
also found $\gamma=16.5$~\eVA{} in a different sample
\cite{jusserand95}. A recent experimental value of
$\gamma=28$~\eVA{} from transport measurements \cite{miller03} is
less direct, includes the cubic Dresselhaus only as a
density-dependent renormalization of the linear Dresselhaus, and
assumes the Rashba coupling is independent of gate voltage.
Theoretical work has indicated that $\gamma$ is smaller in
AlGaAs/GaAs heterostructures and superlattices than it is in bulk
GaAs \cite{santos94,santos95,malcher86}, so it is possible that
experiments are not probing the bulk Dresselhaus coupling, though
Ref.\ \cite{chantis06} predicts $\gamma=8.5$ \eVA{} in bulk GaAs. We
include as supplementary information a table with experimental and
theoretical values of $\gamma$ in GaAs and AlGaAs/GaAs.

Strictly speaking, our calculations are not directly applicable to
the \avg{} data of Zumb\"uhl \etall., as our calculations assume
$E_Z\gg\Delta$, and \avg{} is measured with \Bpar=0. We do not
believe, however, that \apcubic{} changes significantly as
$\Bpar\rightarrow0$; similarly, Cremers \etal consider $a_\Vert$ to
be constant for all \Bpar{} \cite{cremers03}. We believe that
$\apcubic(\Bpar=0)$ can be estimated by simply averaging our results
from the different field directions in the limit
$\Bpar\rightarrow0$.

Our reinterpretation that $a_\Vert^2=\aplinear^2+\apcubic^2$
requires, of course, that \aplinear{} be less than 3.1 in the
experiment of Zumb\"uhl \emph{et al.} This reduction of \aplinear{}
can be absorbed into the geometric parameter $\kappa^\prime$ (which
was set to 1 without fitting in Zumb\"uhl \emph{et al.}) without
affecting any of the physical parameters, $\tau_\phi$,
$\lambda_{SO}$, found by Zumb\"uhl \emph{et al.} Reducing
$\kappa^\prime$ is reasonable, as Ref.~\cite{cremers03} predicts
$\kappa^\prime=1/3$ for a circular diffusive system.

Since $\Delta\propto A^{-1}$, and in a ballistic system $E_T\propto
A^{-1/2}n^{1/2}$, we can see that if the thickness of the 2DEG does
not change with density, $\aplinear\propto A^{7/4}n^{1/4}$, while
$\apcubic\propto\LamD\propto A^{3/4}n^{5/4}$. We therefore expect
that the cubic Dresselhaus spin-orbit coupling should be relatively
more important in small, high density dots, precisely the ones
likely to be useful for producing an SFET.

In summary, we have used billiard simulations to estimate the effect
of the cubic Dresselhaus term on avoided crossings in a closed
chaotic quantum dot. These results are related to the conductance
through a dot with ideal leads attached.  The cubic Dresselhaus
plays a strong and previously ignored role in observed transport
properties through quantum dots. Our calculations suggest that 1)
the Dresselhaus spin-orbit coupling constant, $\gamma$, in
GaAs/AlGaAs heterostructures has a value near 9 \eVA and not the
frequently cited value of 27.5~\eVA, and 2) even with this smaller
value of $\gamma$, in the experiments considered the cubic
Dresselhaus term provided the bulk of the spin-flip portion of the
spin-orbit Hamiltonian, which had previously been assigned to the
effects of linear spin-orbit terms. The value of $\gamma$ in this
technologically important system deserves further study.


%


\begin{figure*}
\includegraphics[width=8in]{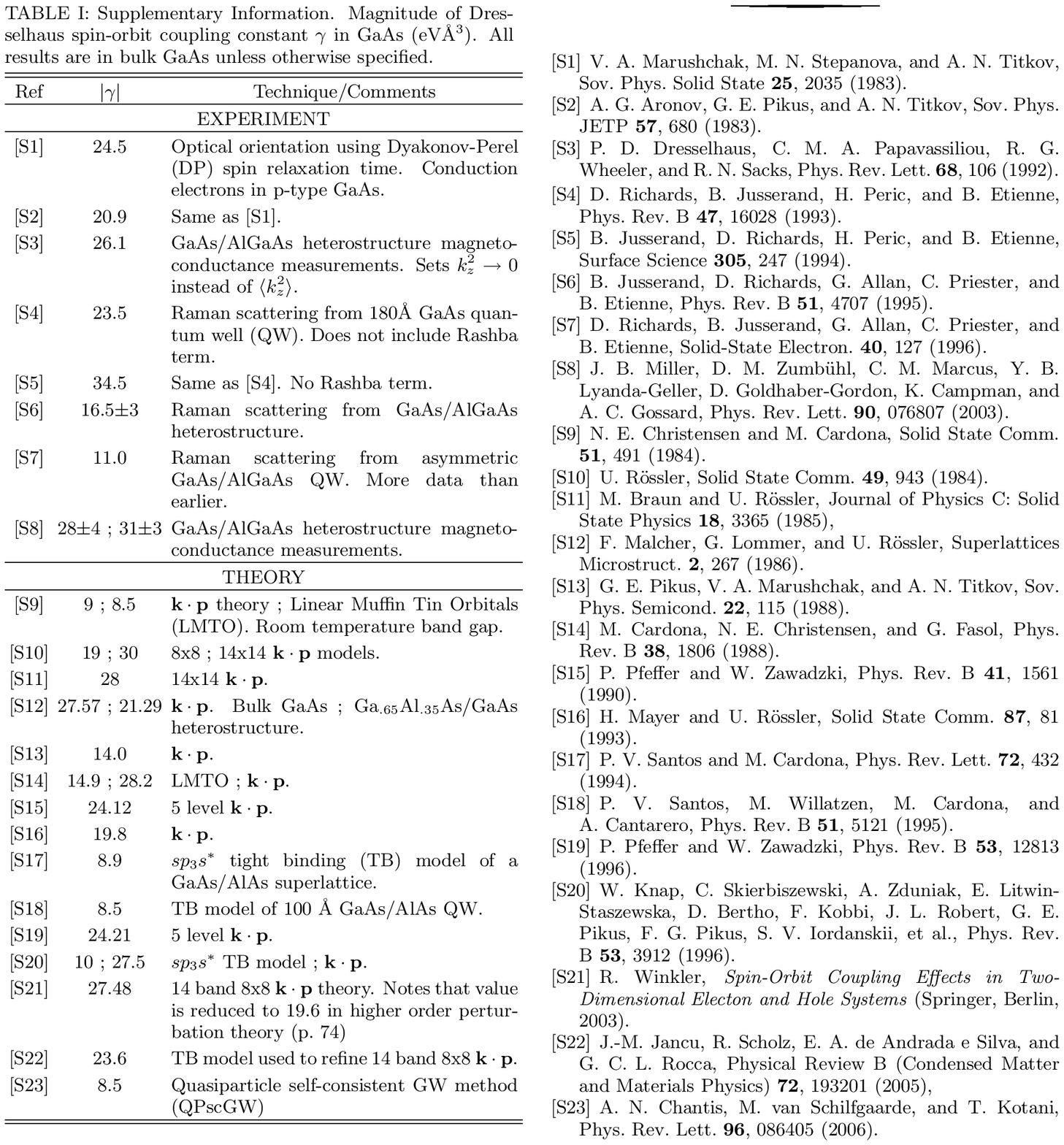}
\end{figure*}

\begin{acknowledgments}
The authors acknowledge a careful reading by Charlie Marcus and
helpful conversations with Dominik Zumb\"uhl, Hakan Tureci, Mike
Stopa, Emmanuel Rashba, Jeff Miller, Subhaneil Lahiri, Eric Heller,
and Hans-Andreas Engel. The work was supported in part by the Fannie
and John Hertz Foundation and NSF grants PHY01-17795 and
DMR05-41988.
\end{acknowledgments}



\bibliography{CubicDressRefs,GammaRefs}


\end{document}